# A Note on Solar Cycle Length during the Medieval Climate Anomaly

José M. Vaquero[1,2] · Ricardo M. Trigo[2,3]

[1] Departamento de Física, Universidad de Extremadura, Mérida, Spain

[2] CGUL-IDL, Universidade de Lisboa, Lisbon, Portugal

[3] Departamento de Eng. Civil da Universidade Lusófona, Lisbon, Portugal

Corresponding author:

José M. Vaquero

Departamento de Física, Centro Universitario de Mérida

Universidad de Extremadura

Avda. Santa Teresa de Jornet, 38

06800 Mérida, Spain

jvaquero@unex.es

**Abstract**

The growing interest in the "Medieval Climate Anomaly" (MCA) and its possible link to anomalous solar activity has prompted new reconstructions of solar activity based on cosmogenic radionuclides. These proxies however do not sufficiently constrain the Total Solar Irradiance (TSI) range, and are often defined at low temporal resolution, inadequate to infer the solar cycle length (SCL). We have reconstructed the SCL (average duration of 10.72 ± 0.20 years) during the MCA using observations of naked-eye sunspot and aurora sightings. Thus, solar activity was most probably not exceptionally intense, supporting the view that internal variability of the coupled ocean–atmosphere system was the main driver of MCA.

**1. Introduction**

The two most important climate events of the last millennium, *i.e.* the "Medieval Climate Anomaly" (MCA, AD 1000–1300) and "Little Ice Age"(LIA, AD 1350–1850), have been linked to relatively prolonged anomalous activity of the Sun, namely the Medieval Solar Maximum (Jirikowic and Damon, 1994) during the 12th and 13th centuries and the Maunder Minimum (Eddy, 1976) between the mid–17th and early–18th centuries.

An increasing number of proxies have become available in the last two decades thus allowing a growing number of more robust paleoclimate reconstructions and modelling studies for the last millennium (Mann *et al.*, 2009). Several possible mechanisms responsible for these two major climatic events have been suggested, including external forcing factors (*e.g.* changes in solar output or volcanism) and internal variability of the coupled ocean-atmosphere system, *e.g.* North Atlantic Oscillation (Trouet *et al.*, 2009) and El Niño–Southern Oscillation (Cobb *et al*., 2003). Nevertheless, solar activity has long been assumed as having played the most relevant role during the late Holocene, particularly in establishing the contrast between the MCA and LIA periods (Díaz *et al*., 2011).

The most common proxies of solar activity are those relative to cosmogenic radionuclides $^{10}$Be and $^{14}$C which are produced by cosmic rays in the Earth's atmosphere (Usoskin, 2008). These proxies however do not sufficiently constrain the Total Solar Irradiance (TSI) range, and additionally are often defined with low temporal resolution (decadal), inadequate to infer the accurate solar-cycle length (SCL). However high-resolution proxy data of solar activity have became available through the compilation of ancient observations of naked-eye sunspot and aurora sightings recorded with fairly precise dates (Vaquero and Vázquez, 2009). Here, we explore this and use the annual number of naked-eye sunspot observations (NE) and the annual number of auroras (NA) observed during the period AD 1000–1300.

**2. MCA and Solar Proxies**

Our knowledge of solar activity around the MCA period (AD 1000–1300), associated with both types of solar proxies is summarized in Figure 1. Black lines show two different reconstructions of the TSI based on $^{10}$Be and $^{14}$C records (Steinhilber *et al*., 2009; Vieira *et al*., 2011). We have used the NE series constructed by Vaquero *et al*. (2002) with updating records from Korean sources (Lee *et al*., 2004). Concerning NA we have used the well-known dataset developed by Křivský and Pejml (1988) that has been updated by the National Geophysical Data Center (NOAA). The total number of NE and NA events accounted for in these two datasets for the period considered in Figure 1 is 50 and 246 respectively. The capacity revealed by both NE and NA series to reproduce the long-term variability of solar activity throughout the entire millennium (Vaquero and Vázquez, 2009) should be stressed.

These solar proxies show that the main phase of the MCA corresponds to a period with moderate-to-high solar magnetic-activity levels in comparison with the modern grand maximum, being delimited by two grand minima episodes, the Oort Minimum (AD 1010–1070) and Wolf Minimum (AD 1270–1340), respectively. Moreover, both reconstructions present four local maxima of TSI during the MCA with an approximate 50-year periodicity. The high-resolution solar proxies NE and NA (blue and orange lines in Figure 1) also show an active period clustered between the Oort and Wolf Minima.

3. Results

We have estimated the timings of 11-year solar cycle maxima using the relative maxima of both proxies (as shown by the blue and orange arrows in Figure 1). We have looked for the local maxima in the NE and NA series to detect maxima of the 11-year solar cycle during the MCA. Although both time series have very high temporal resolution (*i.e.* day, month, year) we have used only the annual frequency of NE and NA to infer the annual level of activity. Using the period 1050–1300, the annual average value of NA and NE are respectively 0.892 (1.287) and 0.199 (0.607) where the number in parentheses indicates the standard deviation. Afterwards, we considered a local maximum of the two series whenever the peak value exceeded the mean value plus three standard deviations, *i.e.* 4.753 and 2.019 for NA and NE respectively. These maxima were dated in the following years: (NE) AD 1139, 1160, and 1185; (NA) AD 1098, 1130, 1138, and 1204. Interestingly, Miyahara *et al*. (2010) estimated independently, based on annual-resolution measurements of a $^{14}$C record from tree rings, the years AD 1098, 1131, and 1140, dates that fall close to our estimations of solar maxima.

We have computed the coefficients of linear relationships between the dates of solar-cycle maxima and the solar cycle number (Figure 2). In this representation the slope value represents the mean value of the solar cycle, having obtained 10.61 ± 0.21, 11.50 ± 0.58 for NA and NE time series, respectively and 10.72 ± 0.20 years when both series are merged. These values represent the estimation of mean SCL for ten consecutive solar cycles (110 years approximately) comprised between 1095 and 1204. Black arrows shown in Figure 1 represent our theoretical ≈11-year solar-cycle maxima. The embedded plot depicts a histogram of the delays (in years) between the mean value from the linear fit and direct measured peak position respectively, i.e. the vertical distance between each point and the fitted line.

## 4. Discussion

We have used the same approach to compute the mean SCL for consecutive sub-periods of ten solar cycles from the late Maunder Minimum until the present (Figure 3, red line) using the well-known Group Sunspot Number (Hoyt and Schatten, 1998; Vaquero, 2007). Results show that the average SCL obtained for the period AD 1095–1204 is similar to the recent period AD 1870–1979.

Note that we are comparing results obtained from no instrumental observation with telescopic observations. It would be very interesting to compare our results directly with the more recent NE observations. Unfortunately NE series are poor after 1600 approximately due to a change in the type of documents used. In any case, modern naked-eye observations of sunspots using filters reproduce clearly the solar cycle (see Chapter 2 in Vaquero and Vázquez, 2009).

Several methods have been proposed to define the SCL (Mursula and Ulich, 1998; Fligge *et al*., 1999; Benestad, 2005; Vaquero *et al*., 2006). Thus, following the more standard procedure, we have also computed the mean SCL using the time intervals from minimum to minimum (Figure 3, blue line). One can confirm from Figure 3 that there is some additional variability of SCL based on the maxima time series (red curve) than when the analysis is performed with the minima (blue curve). Unfortunately, documentary sources such as NA or NE are not sufficiently detailed to define the times of solar cycle minima.

A number of studies in the last decades have focused on the relationships between solar-cycle amplitude and solar-cycle length, *e.g*. Dicke (1978) and Hoyng (1993). Part of this growing interest has been fuelled by the forecasting potential associated with such relationships, particularly taking into account that solar-cycle length precedes the solar-cycle amplitude. For example, Solanki *et al*. (2002) have found a simple empirical relationship between SCL and amplitude using cycle length information of the preceding solar cycles. This empirical model allowed to reproduce the SCA of a given cycle with an average error of 20 in sunspot number. However, the exact relationship might be complex and non-stationary (Solanki *et al*., 2002; Vaquero and Trigo, 2008).

In fact, the correlation between SCA and the preceding SCL is only strongly significant in a statistical sense during the first half of the historical record using the International or Wolf Sunspot Number and it is less significant for the Group Sunspot Number (Vaquero and Trigo, 2008).

Based on maximum (red) and minimum (blue) values of mean SCL represented in Figure 3, it is possible to observe two clusters of low values for sub-periods of ten solar cycles just before the Dalton minimum and during the more recent decades. Moreover, we can state that the average SCL obtained for the period AD 1095–1204 (green line) is similar to those of periods AD 1720–1829 and AD 1870–1979. These periods correspond to time intervals of relatively high secular solar activity. The minimum value of mean SCL for the last three centuries corresponds to the last sub-period (AD 1894–2001) that is considered as the Modern Maximum of solar activity (last value of blue curve), characterised with the highest TSI values for the last millennium (Steinhilber *et al.*, 2009; Vieira *et al.*, 2011).

## 5. Conclusion

In conclusion, high-resolution solar proxies show a very well-defined solar cycle with duration of $10.72 \pm 0.20$ years (average, AD 1095–1204). In this context, we can state that the solar activity during the MCA was not exceptionally anomalous and, most probably neither the corresponding TSI, assuming that the relationship between SCL and SCA established primarily for the recent cycles is universally valid for any other period of time in the past evolution of the solar cycle. This result provides further support for the rationale that either a lack of volcanic activity or the internal variability of the coupled ocean-atmosphere system was the main driver of the MCA. Moreover, contradicting results on the solar-forcing hypothesis have been published in recent years. According to the last IPCC assessment report (Jansen *et al.*, 2007), several model simulations for the last millennium do show warmer temperatures for the MCA (relative to the LIA) when forced with estimated natural forcings. However, a significant number of paleoclimatic modelling efforts do not support the idea that increased solar irradiance implies surface warming in all locations (Ammann *et al.*, 2007; Meehl *et al.*, 2009).

**Acknowledgements**

The authors are grateful to I.G. Usoskin, L.E.A. Vieira, and H. Miyahara for their comments and data. Support from the Junta de Extremadura and Ministerio de Ciencia e Innovación of the Spanish Government (AYA2008-04864/AYA and AYA2011-25945) is gratefully acknowledged. We acknowledge the support from the Portuguese Science Foundation (FCT) through the project MEDIATIC (PTDC/AAC-CLI/103361/2008).

**References**

Ammann, C.M., Joos, F., Schimel, D.S., Otto-Blisner, B.L., Tomas, R.A.: 2007, *Proc. Nat. Acad. Sci.* **104**, 3713.

Benestad, R.E.: 2005, *Geophys. Res. Lett.* **32**, 15714.

Cobb, K.M., Charles, C.D., Cheng, H., Edwards, R.L.: 2003, *Nature* **424**, 271.

Diaz, H.F., Trigo, R.M., Hughes, M.K., Mann, M.E., Xoplaki, E., Barriopedro, D.: 2011, *Bull. Am. Meteor. Soc.* **92**, 1487.

Dicke, R.H.: 1978, *Nature* **276**, 676.

Eddy, J.A.:1976, *Science* **192**, 1189.

Fligge, M., Solanki, S.K., Beer, J.: 1999, *Astron. Astrophys.* **346**, 313.

Hoyng, P.: 1993, *Astron. Astrophys.* **272**, 321.

Hoyt, D.V., Schatten, K.H.: 1998, *Solar Physics* **181**, 491.

Jansen, E., Overpeck, J., Briffa, K.R., Duplessy, J.-C., Joos, F., Masson-Delmotte, V. et al.: 2007, Chapter 6: Paleoclimate. In, Climate Change: The Physical Science Basis. Working Group I Contribution to the Fourth Assessment Report of the Intergovernmental Panel on Climate Change, S. Solomon, D. Qin, M. Manning, Z. Chen, M. Marquis, K.B. Averyt, M. Tignor and H.L. Miller, Eds., Cambridge University Press, New York.

Jirikowic, J.L., Damon, P.E.: 1994, *Clim. Change* **26**, 309.

Křivský. L., Pejml, K.: 1988, Pub. Astron. Inst. Czech. Acad. Sciences, Publication No. 75.

Lee, E.H., Ahn, Y.S., Yang, H.J., Chen, K.Y.: 2004, *Solar Physics* **224**, 373.

Mann, M.E., Zhang, Z., Rutherford, S., Bradley, R.S., Hughes, M.K., Shindell, D., Ammann, C., Falugevi, G., Ni, F.: 2009, *Science* **236**, 1256.

Miyahara, H.: 2010, *J. Geography.* **119**, 510.

Meehl, G.A., Arblaster, J.M., Mathis, K., Sassi, F., van Loon, H.: 2009, *Science* **325**, 1114.

Mursula, K., Ulich, T.: 1998, *Geophys. Res. Lett.* **25**, 1837.

Solanki, S.K., Krivova, N.A., Schüssler, M., Fligge, M.: 2002, *Astron. Astrophys.* **396**, 1029.

Steinhilber, F., Beer, J., Fröhlich, C.: 2009, *Geophys. Res. Lett.* **36**, L19704.

Trouet, V., Esper, J., Graham, N.E., Baker, A., Scourse, J.D., Frank, D.C.: 2009, *Science* **324**, 78.

Usoskin, I.G.: 2008, *Living Rev. Solar Phys.* **5**, 3. [http://www.livingreviews.org/lrsp-2008-3]

Vaquero, J.M.: 2007, *Adv. Space Res.* **40**, 929.

Vaquero, J.M., Gallego, M.C., García, J.A.: 2002, *Geophys. Res. Lett.* **29**, 1997.

Vaquero, J.M., García, J.A., Gallego, M.C.: 2006, *Solar Phys.* **235**, 433.

Vaquero, J.M., Trigo, R.M.: 2008, *Solar Phys.* **250**, 199.

Vaquero, J.M., Vázquez, M.: 2009, The Sun recorded through History, Springer, Dordrecht.

Vieira, L.E.A., Solanki, S.K., Krivova, N.A., Usoskin, I.G.: 2011, *Astron. Astrophys.* **531**, A6.

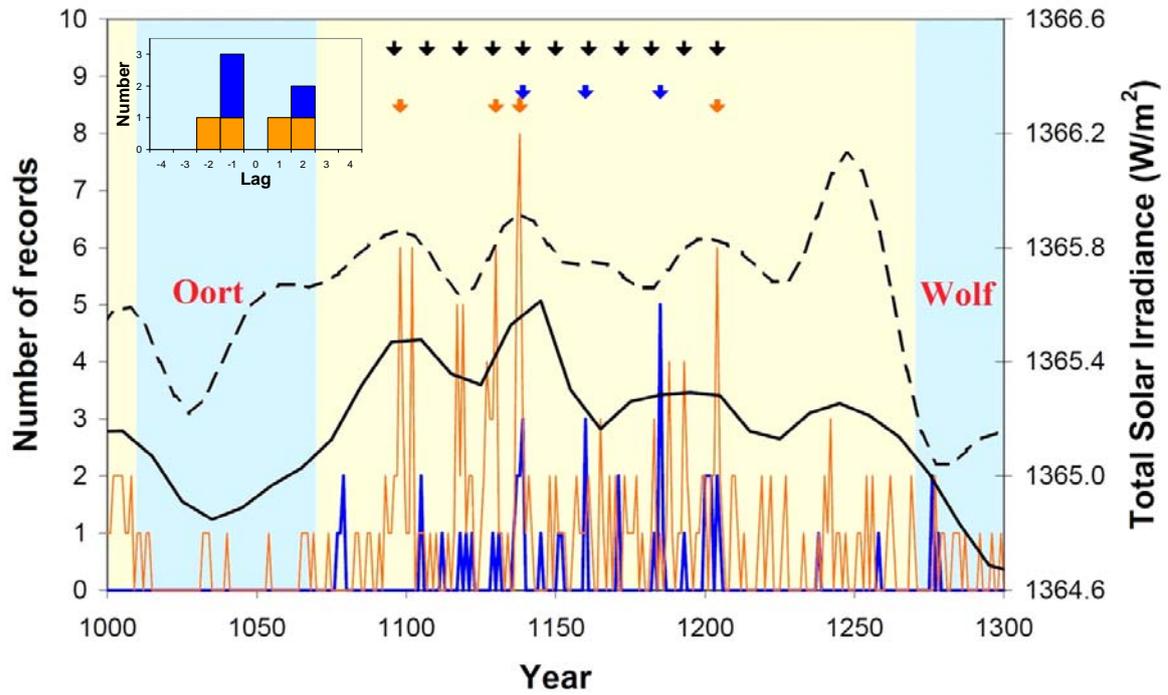

Figure 1. Different solar activity proxies during the period 1050–1300: TSI reconstructed by Steinhilber *et al*. (2009) (dashed black line), TSI reconstructed by Vieira *et al*. (2011) (continuous black line), annual number of naked-eye observations of sunspots (Vaquero *et al*., 2002) (blue line), and annual number of auroral nights (Křivský and Pejml, 1988) (orange line). Black arrows are evenly spaced and correspond to our estimated maxima of solar cycle derived from the linear fit represented in Figure 2. Arrows correspond to estimated maxima of solar cycle using naked-eye observations (blue) and auroral nights (orange). Graphic inserted shows, using the same colour code, a histogram of the delays (in years) between the fitted and estimated maxima of solar cycle.

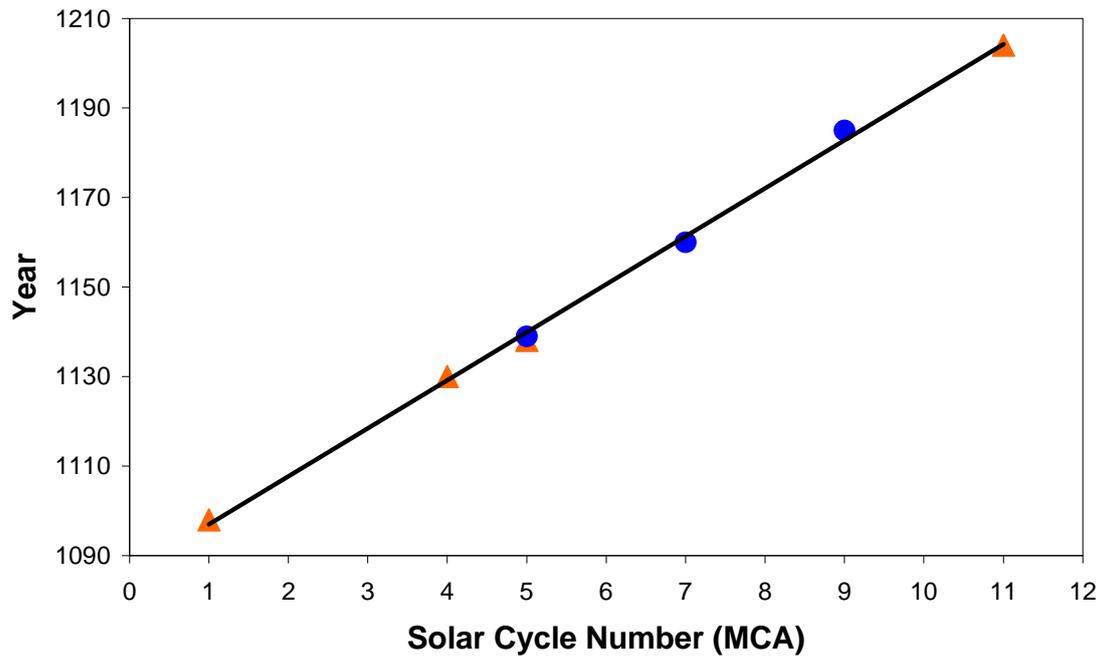

Figure 2. Years of estimated solar-cycle maxima from NA (orange triangles) and NE (blue circles) series *versus* solar cycle number during Medieval Climate Anomaly.

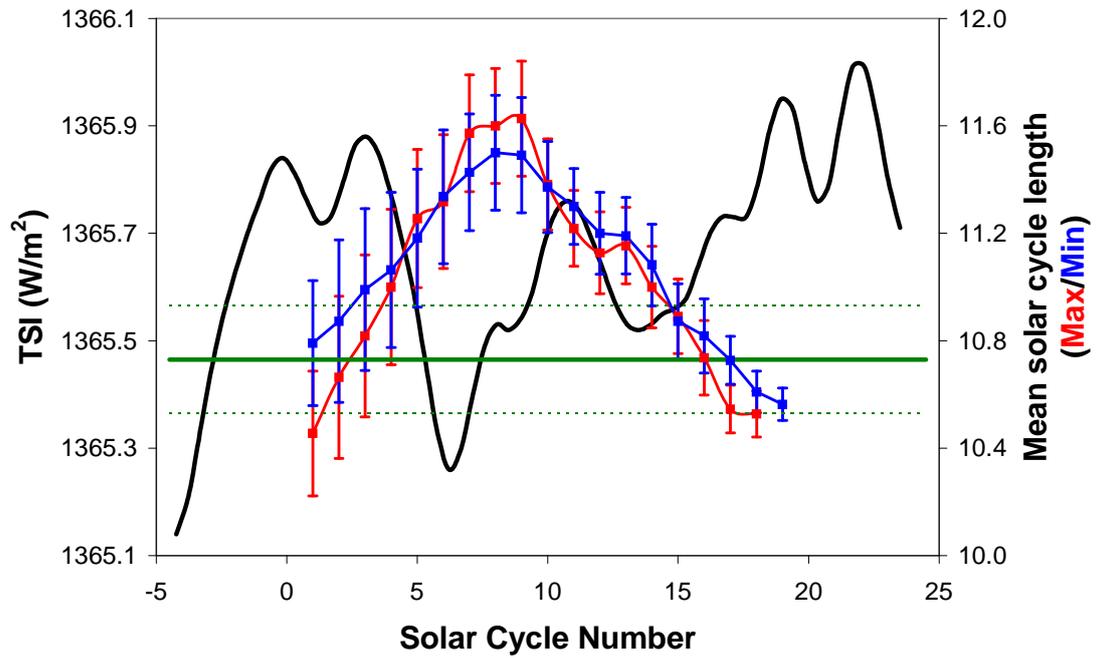

Figure 3. Evolution of mean SCL during the three centuries using (red) max to max or (blue) min to min estimations. Each value represents the estimation of mean SCL for ten consecutive solar cycles (110 years approximately) and the error bars the corresponding standard error. Black line represents the TSI from Steinhilber *et al*. (2009). Green line shows our estimation of mean SCL during MCA including the standard error (dashed lines).